\documentstyle[aps,prb,epsf,twocolumn,floats]{revtex}

\begin{document}
\draft

\title{Theory of directed localization in one dimension}

\author{P.\ W.\ Brouwer, P.\ G.\ Silvestrov,\thanks{On leave of absence from Budker Institute of Nuclear Physics, Novosibirsk, Russia.} and C.\ W.\ J.\ Beenakker}

\address{Instituut-Lorentz, Leiden University, P.O. Box 9506, 2300 RA
Leiden, The Netherlands\\
{\rm (\today)}
\medskip ~~\\ \parbox{14cm}{\rm
We present an analytical solution of the delocalization transition that is induced by an imaginary vector potential in a disordered chain [N.\ Hatano and D.\ R.\ Nelson, Phys.\ Rev.\ Lett.\ {\bf 77}, 570 (1996)]. We compute the relation between the real and imaginary parts of the energy in the thermodynamic limit, as well as finite-size effects. The results are in good agreement with numerical simulations for weak disorder (mean free path large compared to the wavelength). 
\vspace{-0.05\hsize}
}}


\maketitle
\narrowtext

In a recent Letter,\cite{HatanoNelson} Hatano and Nelson have demonstrated the existence of a mobility edge in a disordered ring with an imaginary vector potential. A non-Hermitian Hamiltonian containing an imaginary vector potential arises from the study of the pinning of vortices by columnar defects in a superconducting cylinder.\cite{NelsonVinokur} Their discovery of a delocalization transition in one and two-dimensional systems has generated considerable interest,\cite{Efetov,FeinbergZee,Chalker} since all states are localized by disorder in one and two dimensions if the vector potential is real. Localization in this specific kind of non-Hermitian quantum mechanics is referred to as ``directed localization'',\cite{Efetov} because the imaginary vector potential breaks the symmetry between left-moving and right-moving particles, without breaking time-reversal symmetry.

The analytical results of Ref.\ \onlinecite{HatanoNelson} consist of an expression for the mobility edge plus a solution of the one-dimensional problem with a single impurity. Here we go further, by solving the many-impurity case in one dimension. Most of the technical results which we will need were derived previously in connection with the problem of localization in the presence of an imaginary {\em scalar} potential. Physically, these two problems are entirely different: an imaginary vector potential singles out a direction in space, while an imaginary scalar potential singles out a direction in time: A negative imaginary part of the scalar potential corresponds to absorption and a positive imaginary part to amplification. One might surmise that amplification could cause a delocalization transition, but in fact all states remain localized in one dimension in the presence of an imaginary scalar potential.\cite{Zhang,Paasschens}

Following Ref.\ \onlinecite{HatanoNelson} we consider a disordered chain with single-particle Hamiltonian
\begin{equation}
  {\cal H} = -{w \over 2} \sum_{j} \left(e^{ha} c^{\dagger}_{j+1} c^{\vphantom{\dagger}}_{j} + e^{-ha} c^{\dagger}_{j} c^{\vphantom{\dagger}}_{j+1} \right) + \sum_{j} V_{j}^{\vphantom{\dagger}} c^{\dagger}_{j} c^{\vphantom{\dagger}}_{j}. \label{eq:Hlattice}
\end{equation}
The operators $c^{\dagger}_{j}$ and $c^{\vphantom{\dagger}}_{j}$ are creation and annihilation operators, $a$ is the lattice constant, and $w$ the hopping parameter. The random potential $V_{j}$ is chosen independently for each site, from a distribution with zero mean and variance $u^2$. For weak disorder (mean free path much larger than the wavelength), higher moments of the distribution of $V_{j}$ are not relevant. The Hamiltonian is non-Hermitian because of the real parameter $h$, corresponding to the imaginary vector potential. The chain of length $L$ is closed into a ring, and the problem is to determine the eigenvalues $\varepsilon$ of ${\cal H}$. If $\varepsilon$ is an eigenvalue of ${\cal H}$, than also $\varepsilon^{*}$ is one --- because ${\cal H}$ is real. Real $\varepsilon$ correspond to localized states, while complex $\varepsilon$ correspond to extended states.\cite{HatanoNelson}

To solve this problem, we reformulate it in terms of the $2 \times 2$ transfer matrix $M_{h}(\varepsilon)$ of the chain, which relates wave amplitudes at both ends.\cite{DorokhovAB} The energy $\varepsilon$ is an eigenvalue of ${\cal H}$ if and only if $M_{h}(\varepsilon)$ has an eigenvalue $1$. The use of the transfer matrix is advantageous, because the effect of the imaginary vector potential is just to multiply $M$ with a scalar,
\begin{equation}
  M_h(\varepsilon) = e^{h L} M_0(\varepsilon). \label{eq:Mh}
\end{equation}
The energy spectrum is therefore determined by
\begin{equation}
  \det \left[1 - e^{h L} M_0(\varepsilon) \right] = 0. \label{eq:detM}
\end{equation}
Time-reversal symmetry implies $\det M_0 = 1$. Hence the determinantal equation (\ref{eq:detM}) is equivalent to\cite{footnoteHJT}
\begin{equation}
  \mbox{tr}\, M_0(\varepsilon) = 2 \cosh hL. \label{eq:Mcond}
\end{equation}
We seek the solution in the limit $L \to \infty$.

Since $M_0$ is the transfer matrix in the absence of the imaginary vector potential ($h=0$), we can use the results in the literature on localization in conventional one-dimensional systems (having an Hermitian Hamiltonian).\cite{Pendry} The four matrix elements of $M_0$ are given in terms of the reflection amplitudes $r$, $r'$ and the transmission amplitude $t$ by
\begin{equation}
  \begin{array}{ll}
    (M_0)_{11} = -(1/t) \det S, & (M_0)_{12} = r'/t, \\ 
    (M_0)_{21} = -r/t, & (M_0)_{22} = 1/t,
  \end{array}
\end{equation}
where $\det S = r r' - t^2$ is the determinant of the scattering matrix.
(There is only a single transmission amplitude because of time-reversal symmetry, so that transmission from left to right is equivalent to transmission from right to left.) The transmission probability $T = |t|^2$ decays exponentially in the large-$L$ limit, with decay length $\xi$:
\begin{equation}
  - \lim_{L \to \infty} L^{-1} \ln T = \xi^{-1}. \label{eq:TL}
\end{equation}
The energy dependence of $\xi$ is known for weak disorder, such that $|k|\xi \gg 1$, where the complex wavenumber $k$ is related to $\varepsilon$ by the dispersion relation
\begin{equation}
  \varepsilon = -w \cos ka. \label{eq:dispersion}
\end{equation}
For real $k$, the decay length is the localization length $\xi_0$, given by \cite{DorokhovPRB}
\begin{equation}
  \xi_0 = a (w/u)^2 \sin^2 (\mbox{Re}\, ka). \label{eq:xi0}
\end{equation}
(Since $\xi_0$ is of the order of the mean free path $\ell$, the condition of weak disorder requires $\ell$ large compared to the wavelength.)
For complex $k$, the decay length is shorter than $\xi_0$, regardless of the sign of $\mbox{Im}\, k$, according to \cite{Paasschens}
\begin{equation}
  \xi^{-1} = \xi_0^{-1} + 2 |\mbox{Im}\, k|. \label{eq:xicomplex}
\end{equation}

We use these results to simplify Eq.\ (\ref{eq:Mcond}). Upon taking the logarithm of both sides of Eq.\ (\ref{eq:Mcond}), dividing by $L$ and taking the limit $L \to \infty$, one finds
\begin{equation}
  |h| - \case{1}{2}\xi^{-1} = \lim_{L \to \infty} L^{-1} \ln|1 - \det S|, \label{eq:trM}
\end{equation}
where we have used that $L^{-1} \ln f \to L^{-1} \ln |f|$ as $L \to \infty$ for any complex function $f(L)$. For complex $k$, the absolute value of $\det S$ is either $< 1$ (for $\mbox{Im}\, k > 0$) or $> 1$ (for $\mbox{Im}\, k < 0$). As a consequence, $\ln|1-\det S|$ remains bounded for $L \to \infty$, so that the right-hand-side of Eq.\ (\ref{eq:trM}) vanishes. Substituting Eq.\ (\ref{eq:xicomplex}), we find that complex wavenumbers $k$ satisfy
\begin{equation}
  |\mbox{Im}\, k| = |h| - \case{1}{2} \xi_0^{-1}. \label{eq:Imk}
\end{equation}
Together with the expression (\ref{eq:xi0}) for the localization length $\xi_0$, this is a relation between the real and imaginary parts of the wavenumber. Using the dispersion relation (\ref{eq:dispersion}), and noticing that the condition $|k| \xi \gg 1$ for weak disorder implies $|\mbox{Im}\, k| \ll |\mbox{Re}\, k|$, we can transform Eq.\ (\ref{eq:Imk}) into a relation between the real and imaginary parts of the energy,
\begin{equation}
  |\mbox{Im}\, \varepsilon| = 
|h| a \sqrt{ w^2 - (\mbox{Re}\, \varepsilon)^2 } - {u^2 \over 2 \sqrt{ w^2 - (\mbox{Re}\, \varepsilon)^2}}.
  \label{eq:spectrum}
\end{equation}

The support of the density of states in the complex plane consists of the closed curve (\ref{eq:spectrum}) plus two line segments on the real axis,\cite{footnote} extending from the band edge $\pm w$ to the mobility edge $\pm \varepsilon_c$. The real eigenvalues are identical to the eigenvalues at $h=0$, up to exponentially small corrections. The energy $\varepsilon_c$ is obtained by putting $\mbox{Im}\, \varepsilon = 0$ in Eq.\ (\ref{eq:spectrum}), or equivalently be equating\cite{HatanoNelson} $2 \xi_0$ to $1/|h|$, hence
\begin{equation}
  \varepsilon_c = (w^2 - u^2/2 |h| a)^{1/2}.
\end{equation}
The delocalization transition at $\varepsilon_c$ exists for $|h| > h_c = \case{1}{2} u^2/w^2 a$. 
\begin{figure}
\epsfxsize=0.89\hsize
\hspace{0.06\hsize}
\epsffile{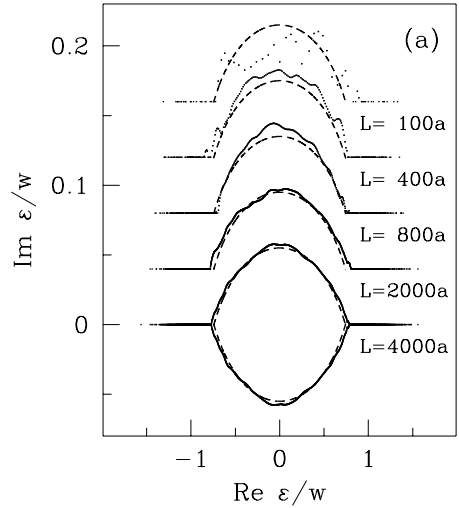}
\vspace{-0.1\hsize}

\epsfxsize=0.89\hsize
\hspace{0.06\hsize}
\epsffile{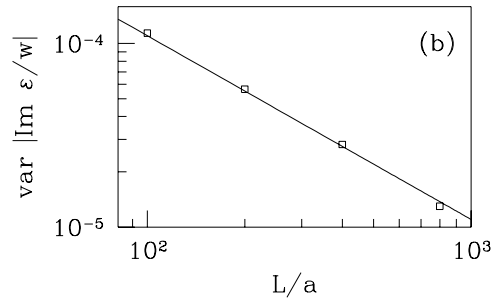}
\vspace{-0.1\hsize}

\caption{\label{fig:1} (a) Data points: eigenvalues of the Hamiltonian (\protect\ref{eq:Hlattice}), for parameter values $ha = 0.1$, $u/w = 0.3$, and for five values of $L/a$. Dashed curves: analytical large-$L$ limit, given by Eq.\ (\protect\ref{eq:spectrum}). (Except for the case $L=4000a$, spectra are offset vertically and only eigenvalues with $\mbox{Im}\, \varepsilon \ge 0$ are shown.) (b) Variance of the imaginary part of the eigenvalues as a function of the sample length, for $\mbox{Re}\, \varepsilon \approx 0$ and for the same parameter values as in (a). The data points are the numerical results for 1000 samples. The solid line is the analytical result (\protect\ref{eq:var}).}
\end{figure}

In Fig.\ \ref{fig:1}a, the analytical theory is compared with a numerical diagonalization of the Hamiltonian (\ref{eq:Hlattice}). The numerical finite-$L$ results are consistent with the large-$L$ limit (dashed curve). To leading order in $1/L$, fluctuations of $\mbox{Im}\, \varepsilon$ around the large-$L$ limit (\ref{eq:spectrum}) are governed by fluctuations of the transmission probability $T$. (Fluctuations of $L^{-1} \ln T$ are of order $L^{-1/2}$, while the other fluctuating contributions to Eq.\ (\ref{eq:Mcond}) are of order $L^{-1}$.) The variance of $\ln T$ for large $L$ is known,\cite{Misirpashaev} 
\begin{equation}
  \mbox{var}\, \ln T = {2L \over \xi_0} + 8L |\mbox{Im}\, k| e^{4 \xi_0 |{\rm Im}\, k|} \mbox{Ei}\,(-4 \xi_0 |\mbox{Im}\, k|),
\end{equation}
where $\mbox{Ei}$ is the exponential integral. Equating $|\mbox{Im}\, k| = |h| + \case{1}{2} L^{-1} \ln T$, we find $\mbox{var}\, |\mbox{Im}\, k| = \case{1}{4} L^{-2}\, \mbox{var}\, \ln T$ and thus
\begin{equation}
  \mbox{var}\, |\mbox{Im}\, \varepsilon| = {a^2 \over 2 L \xi_0} 
    \left[ {1 + 2 \gamma e^{2 \gamma} \mbox{Ei}\,(-2 \gamma)} \right] \left[ {w^2 - (\mbox{Re}\, \varepsilon)^2} \right], \label{eq:var}
\end{equation}
where $\gamma = 2 |h| \xi_0 - 1$. In Fig.\ \ref{fig:1}b we see that Eq.\ (\ref{eq:var}) agrees well with the results of the numerical diagonalization.
The fluctuations $\Delta\, \mbox{Im}\, \varepsilon$ are correlated over a range $\Delta \mbox{Re}\, \varepsilon$ which is large compared to $\Delta \mbox{Im}\, \varepsilon$ itself,\cite{footnoteP} their ratio $\Delta\, \mbox{Im}\, \varepsilon/\Delta\, \mbox{Re}\, \varepsilon$ decreasing $\propto L^{-1/2}$. This explains why the complex eigenvalues for a specific sample appear to lie on a smooth curve (see Fig.\ \ref{fig:1}a). This curve is sample-specific and fluctuates around the large-$L$ limit (\ref{eq:spectrum}). 

In conclusion, we have presented an analytical theory for the delocalization transition in a single-channel disordered wire with an imaginary vector potential. We find good agreement with numerical diagonalizations, both for the relation between the real and imaginary parts of the energy in the infinite-length limit and for the finite-size effects. In the absence of the imaginary vector potential, the transfer matrix approach used in this paper has been very successful for the study of localization in disordered wires with more than one scattering channel. We expect that such an extension of the theory is possible for non-Hermitian systems as well.

Discussions with K.\ B.\ Efetov and D.\ R.\ Nelson motivated us to do this work. We have also benefitted from discussions with T.\ Sh.\ Misirpashaev. Support by the Dutch Science Foundation NWO/FOM is gratefully acknowledged.

{\em Note added:} Just before submitting our paper, we learned of a different analytical approach to this problem by Janik et al.\cite{Janik}


\begin{references}
\bibitem{footnote} One might wonder whether the line segments at real $\varepsilon$ could extend into the interior of the curve (\ref{eq:spectrum}) of complex $\varepsilon$. This is not possible in the large-$L$ limit since $|h| - 1/2 \xi_0 > 0$ for complex $\varepsilon$ [see Eq.\ (\ref{eq:Imk})], while real eigenvalues of ${\cal H}$ appear only if $|h| - 1/2 \xi_0 < 0$. This last inequality follows from Eq.\ (\ref{eq:trM}), which for real $\varepsilon$ equates $|h| - 1/2 \xi_0$ to a quantity $L^{-1} \ln|1 - \det S|$ which oscillates between $-\infty$ and $0$ as $L \to \infty$.
\bibitem{footnoteP} An estimate for $(\Delta\, \mbox{Im}\, \varepsilon/\Delta\, \mbox{Re}\, \varepsilon)^2$ follows from the variance of the slope $d\, \mbox{Im}\, \varepsilon/d\, \mbox{Re}\, \varepsilon$, which for large $L$ is given by
$$
  \mbox{var}\, {d\, \mbox{Im}\, \varepsilon \over d\, \mbox{Re}\, \varepsilon} = {a \over 2 \pi^2 L\, |\mbox{Im}\, \varepsilon|} \sqrt{w^2 - (\mbox{Re}\, \varepsilon)^2}.
$$
\end{references}
\end{document}